\documentclass[pdflatex,sn-mathphys-num]{sn-jnl}

\usepackage{aas_macros}
\usepackage{graphicx}%
\usepackage{multirow}%
\usepackage{amsmath,amssymb,amsfonts}%
\usepackage{amsthm}%
\usepackage{mathrsfs}%
\usepackage[title]{appendix}%
\usepackage{xcolor}%
\usepackage{textcomp}%
\usepackage{manyfoot}%
\usepackage{booktabs}%
\usepackage{algorithm}%
\usepackage{algorithmicx}%
\usepackage{algpseudocode}%
\usepackage{listings}%
\usepackage{siunitx}

\usepackage{upgreek}
\usepackage[T1]{fontenc} 

\newcommand{\um}{$\upmu\mathrm{m}$}
\graphicspath{{./}{figures/}}

\begin{document}

\title{Cross-Calibration of Chandrayaan-2 XSM with INSPIRESat-1 DAXSS and GOES-16 XRS}

\author*[1]{\fnm{N.~P.~S.}~\sur{Mithun}\href{https://orcid.org/0000-0003-3431-6110}{\includegraphics[scale=1.0]{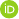}}}\email{mithun@prl.res.in}
\author[1]{\fnm{Santosh V.}~\sur{Vadawale}\href{https://orcid.org/0000-0002-2050-0913}{\includegraphics[scale=1.0]{orcid.pdf}}}
\author[2]{\fnm{Asif~M.}~\sur{Mandayapuram}\href{https://orcid.org/0000-0003-4442-5994}{\includegraphics[scale=1.0]{orcid.pdf}}}
\author[1]{\fnm{M.}~\sur{Shanmugam}\href{https://orcid.org/0000-0002-5995-8681}{\includegraphics[scale=1.0]{orcid.pdf}}}
\author[1]{\fnm{Soumya}~\sur{Kohli}\href{https://orcid.org/0009-0002-9416-2184}{\includegraphics[scale=1.0]{orcid.pdf}}}
\author[2]{\fnm{Mehul R.}~\sur{Pandya}\href{https://orcid.org/0000-0002-3564-7694}{\includegraphics[scale=1.0]{orcid.pdf}}}
\author[1]{\fnm{Anil}~\sur{Bhardwaj}\href{https://orcid.org/0000-0003-1693-453X}{\includegraphics[scale=1.0]{orcid.pdf}}}
\author[3]{\fnm{Robert}~\sur{Sewell}\href{https://orcid.org/0000-0002-3440-2492}{\includegraphics[scale=1.0]{orcid.pdf}}}
\author[4]{\fnm{Bennet}~\sur{Schwab}\href{https://orcid.org/0000-0002-1426-6913}{\includegraphics[scale=1.0]{orcid.pdf}}}
\author[5]{\fnm{Amir}~\sur{Caspi}\href{https://orcid.org/0000-0001-8702-8273}{\includegraphics[scale=1.0]{orcid.pdf}}}
\author[3]{\fnm{Thomas N.}~\sur{Woods}\href{https://orcid.org/0000-0002-2308-6797}{\includegraphics[scale=1.0]{orcid.pdf}}}
\author[6,7]{\fnm{Janet L.}~\sur{Machol}\href{https://orcid.org/0000-0002-0344-0314}{\includegraphics[scale=1.0]{orcid.pdf}}}

\affil[1]{Physical Research Laboratory, Navrangpura, Ahmedabad, Gujarat 380009, India}
\affil[2]{Space Applications Centre, Indian Space Research Organisation, Ahmedabad, Gujarat 380015, India}
\affil[3]{Laboratory for Atmospheric and Space Physics, University of Colorado Boulder, 3665 Discovery Dr., Boulder, CO 80303, USA}
\affil[4]{Space Sciences Laboratory, University of California Berkeley, 7 Gauss Way, Berkeley, CA 94720, USA}
\affil[5]{Southwest Research Institute, 1301 Walnut St., Suite 400, Boulder, CO 80302, USA}
\affil[6]{Cooperative Institute for Research in Environmental Sciences, University of Colorado Boulder, Boulder, CO, USA}
\affil[7]{NOAA National Centers for Environmental Information, Boulder, CO, USA}

\abstract{
X-ray spectroscopic observations of the solar corona and flares provide crucial diagnostics of plasma properties and are essential for understanding the physical processes responsible for coronal heating and solar eruptive activity.
The Chandrayaan-2 Solar X-ray Monitor (XSM) provides disk-integrated spectra of the Sun in the 1--15~keV soft X-ray band, enabling modeling of the thermal X-ray emission from the corona across quiet phases to intense solar flares. XSM has been operational for about seven years, starting from the last solar minimum and covering the maximum of the current Solar Cycle. The Dual-zone Aperture X-ray Solar Spectrometer (DAXSS) instrument on board INSPIRESat-1 covers the solar X-ray spectra in a similar energy range as XSM and was operational during 2022--2026. With multiple instruments simultaneously observing the Sun in X-rays, there is scope to compare measurements across instruments. 
Here, we present the cross-calibration of XSM with DAXSS and a broadband X-ray flux monitor, the GOES-16 X-ray Sensor (XRS). 
Comparisons of XSM and DAXSS spectra reveal an unaccounted attenuation in the XSM low-energy response. Supported by laboratory measurements, we attribute this difference to the effective detector beryllium window thickness being \SI{25}{\micro\meter} rather than the previously assumed \SI{8}{\micro\meter}.
Incorporating this revision into the XSM calibration significantly improves the agreement between the two instruments, with flux measurements agreeing within $\sim$10\% in the 1--8~{\AA} band. Comparison with GOES-16 XRS measurements over a broad range of solar activity levels further demonstrates consistency, with a median flux difference of less than 10\%. 
}

\keywords{Solar X-ray Spectroscopy, Calibration, Solar Flares}
\maketitle
\section{Introduction}
\label{sec:intro}

X-ray emission provides a direct diagnostic of the plasma conditions during solar flares, allowing measurements of temperature, emission measure, and nonthermal electron distributions \citep[e.g.,][]{2018LRSP...15....5D}. Soft X-rays, typically below $\sim$10~keV, primarily originate from thermal continuum processes and line emission, while hard X-rays arise from non-thermal bremsstrahlung emission from accelerated electrons. 

X-Ray Sensor (XRS) instruments on the GOES series of satellites have been used over the past five decades to monitor the broadband solar flux in soft X-rays, and the X-ray flux in 1--8~{\AA} band (1.55--12.4~keV) is used to classify solar flare activity~\citep{1994SoPh..154..275G,2009SPIE.7438E..02C,2024JGRA..12932925W,2026JGRA..13135181M}. While broadband flux measurements are extremely useful, especially given the availability of data sets spanning several solar cycles, they have limited capacity to provide detailed diagnostics of the solar corona and probe the underlying physics. 

X-ray spectroscopic measurements, ideally spanning broader energy bands, are extremely useful for such studies. The Reuven Ramaty High Energy Solar Spectroscopic Imager \citep[RHESSI;][]{2002SoPh..210....3L}, with its X-ray imaging and hard X-ray spectroscopic capabilities, provided a wealth of knowledge on flare physics. However, RHESSI measurements were limited to greater than $\sim$3~keV and could not probe the thermal emission in soft X-rays below this range. Subsequent experiments, such as the GSAT-2 Solar X-ray Spectrometer \citep[SOXS;][]{2005SoPh..227...89J} and the CORONAS-PHOTON Solar Photometer in X-rays \citep[SphinX;][]{2013SoPh..283..631G}, demonstrated the feasibility of extending the energy range down to soft X-rays. The Miniature X-ray Solar Spectrometer \citep[MinXSS;][]{2016JSpRo..53..328M} employed silicon drift detectors (SDD) that provide superior spectral resolution, demonstrating the capabilities of such measurements~\citep{caspi2015,moore18}. 

Over the past few years, several X-ray spectrometers have become operational, capable of spectral measurements across different energy bands from soft to hard X-rays.  
These include Chandrayaan-2 Solar X-ray Monitor \citep[XSM;][]{shanmugam20,2020SoPh..295..139M} in 1--15~keV, Solar Orbiter Spectrometer/Telescope for Imaging X-rays \citep[STIX;][]{2020A&A...642A..15K} in 4--150~keV, INSPIRESat-1 Dual-zone Aperture X-ray Solar Spectrometer \citep[DAXSS;][]{2023ApJ...956...94W} in 0.5--15~keV,  ASO-S Hard X-ray Imager \citep[HXI;][]{2019RAA....19..160Z} in 30--200~keV, and most recently, Aditya-L1 Solar Low Energy X-ray Spectrometer \citep[SoLEXS;][]{2025SoPh..300...87S} in 2--15~keV and High Energy L1 Orbiting X-ray Spectrometer \citep[HEL1OS;][]{2025SoPh..300..140N} in 8--150~keV. 

With several solar X-ray spectroscopic instruments operating simultaneously and overlapping in energy bands, there are opportunities for comparisons and cross-calibrations that enable us to assess the consistency of flux measurements and derived spectral parameters.
While astrophysical X-ray missions often rely on standard candles such as the Crab pulsar and nebula~\citep{2005SPIE.5898...22K}, which has a featureless power-law spectrum with reasonably stable spectral parameters, the effective areas of most solar X-ray spectrometers are very low, making calibration with the Crab extremely difficult. On the other hand, the solar X-ray spectrum shows several soft X-ray spectral lines in addition to the continuum and is highly variable. Thus, truly simultaneous observations of the Sun by multiple instruments provide the most reliable means of cross-calibration \citep[e.g.,][]{2025SoPh..300...56L}.  

Chandrayaan-2, the second lunar mission of the Indian Space Research Organization (ISRO), included the Solar X-ray Monitor \citep[XSM;][]{shanmugam20,2020SoPh..295..139M} with an objective to support the global elemental abundance mapping of the Moon with a remote X-ray fluorescence spectroscopy instrument CLASS~\citep{radhakrishna20}. XSM employs an SDD and provides disk-integrated soft X-ray spectra over the energy range of 1--15~keV, with a 1-second cadence and a spectral resolution of 175~eV FWHM at 5.9~keV. 
The instrument is optimized to handle a large dynamic range, from sub-A-class level activity to X9-class flares, without saturation~\citep{2020SoPh..295..139M}. Various aspects of the XSM spectral response have been calibrated on the ground~\citep{mithun20_gcal} and refined further with on-board measurements~\citep{2020SoPh..295..139M}, and spectroscopy above 1.3~keV has been used for various studies \citep[e.g.,][]{2021ApJ...912L..13V,2021ApJ...920....4M,2022ApJ...939..112M}. However, the spectral response below 1.3~keV was considered uncertain and was not recommended for analysis~\citep{2020SoPh..295..139M}. XSM began observations in September~2019 and continues to operate nominally.

DAXSS instrument on INSPIRESat-1 employs an SDD similar to XSM, covering a similar energy range with a resolution of $\sim$135~eV FWHM at 5.9~keV. By employing a dual-zone aperture design with an annular Kapton filter in front of the detector, DAXSS has a higher effective area above $\sim$1.5~keV than XSM~\citep{2020ApJ...904...20S}; however, this limits the instrument's dynamic range to about the M2 class of flares~\citep{2023ApJ...956...94W}. DAXSS began observations soon after the launch of INSPIRESat-1 in February~2022, and the satellite recently had its atmospheric re-entry in May~2026. Although only a fraction of the DAXSS data could be downloaded due to limited communication links~\citep{2023ApJ...956...94W}, there is some overlap with XSM observations, enabling cross-calibration of the two instruments. The XRS instrument of the GOES-16 satellite~\citep{2024JGRA..12932925W, 2026JGRA..13135181M}, providing broadband X-ray flux in two bands of 0.5--4~{\AA} and 1--8~{\AA}, was also operational simultaneously to XSM. Although spectral comparisons similar to DAXSS are not feasible in the case of GOES-16 XRS, flux comparisons over longer observation periods are possible. 

Here, we present the cross-calibration of the Chandrayaan-2 XSM with the INSPIRESat-1 DAXSS and the GOES-16 XRS instruments. Details of the observations and data reduction are presented in Section~\ref{sec:obs}. Results of the spectral and flux comparisons of XSM with DAXSS, and the flux comparison of XSM with GOES XRS are discussed in Section~\ref{sec:result}, followed by a summary in Section~\ref{sec:summary}.

\section{Observations and Data Reduction}
\label{sec:obs}

\begin{figure*}
\begin{center}
   \includegraphics[width=0.99\textwidth]{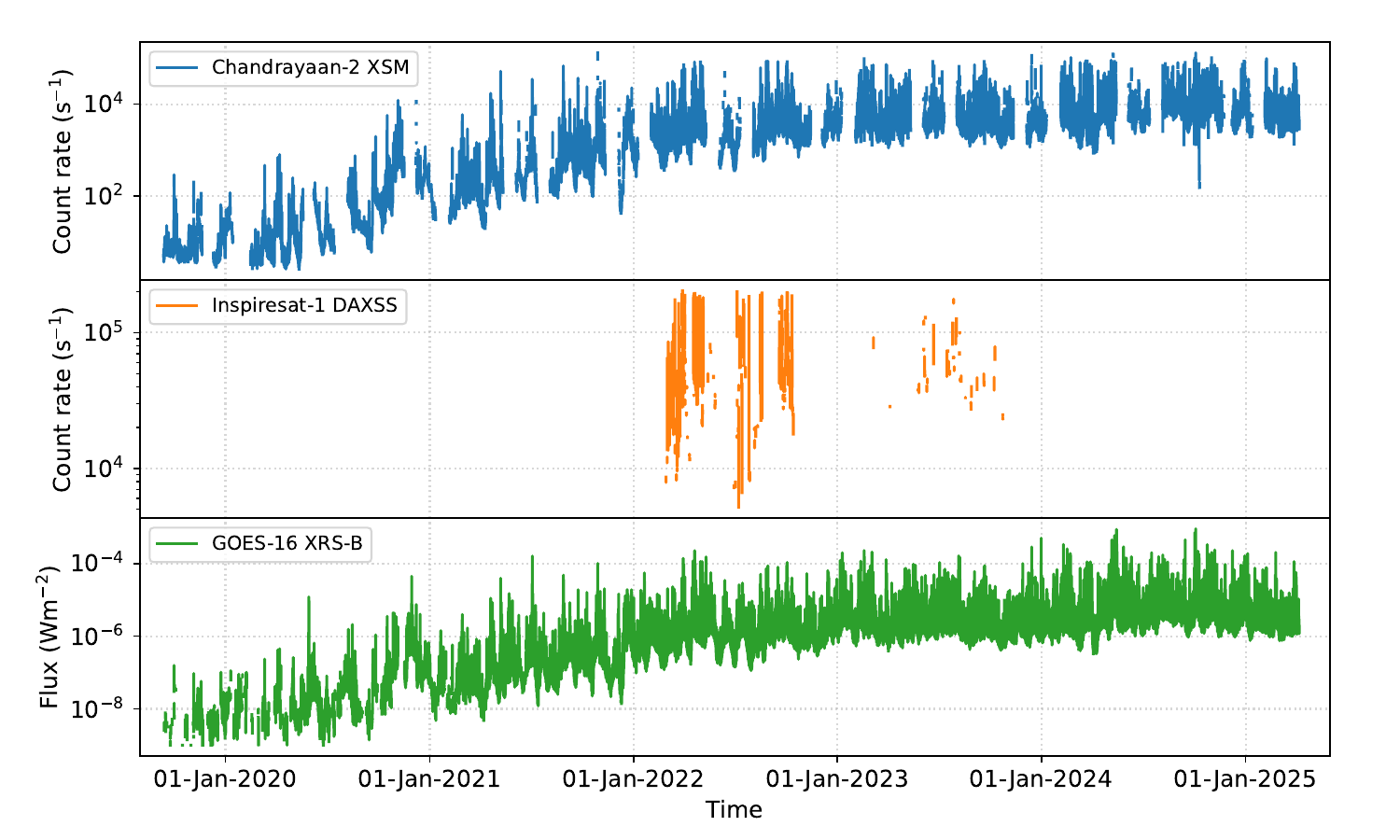}
   \caption{Light curves observed with Chandrayaan-2 XSM, INSPIRESat-1 DAXSS, and GOES-16 XRS `B' channel during the overlapping periods of observations of the three instruments. The time cadence of XSM and DAXSS count rates is 1~minute and 54~seconds, respectively. For XRS, the 1-minute average 1--8~{\AA} band (Channel B) flux is shown. \label{lcfig}}
\end{center}
\end{figure*}

Solar X-ray spectral data with XSM are available at a 1-second cadence from 12~September~2019 and have good temporal coverage during alternating approximately three-month-long dawn-dusk seasons~\citep{2020SoPh..295..139M}. DAXSS spectral data are available at a 9-second cadence for selected periods of flares and quiet phases between February~2022 and October~2023. 
DAXSS data storage had stopped in late~2023 due to INSPIRESat-1 SD-card failures. With reconfiguration to use the INSPIRESat-1 flash memory, limited additional solar data acquisitions with DAXSS have been made during 2025--2026, and these data products are expected to become available in fall~2026.
X-ray flux estimates in two wavelength bands, 1--8~{\AA} and 0.5--4~{\AA}, have been measured at 1-second cadence by GOES-16 XRS continuously from February~2017 (before the start of XSM operations) until April~2025 (GOES-18 is the operational satellite after this). Figure~\ref{lcfig} shows the X-ray light curves from XSM, DAXSS, and GOES-16 XRS during the overlapping periods of observations of the three instruments.        

For comparisons of DAXSS spectra with XSM, we use the DAXSS Level-1 data V3.1.0\footnote{\url{https://lasp.colorado.edu/minxss/data/}}. Although the DAXSS spectra are available at 9-s cadence, there are insufficient statistics in both instruments within this native time bin. We therefore generate DAXSS spectra in 54-s time bins by summing the spectra from consecutive 9-s time bins, ignoring cases where gaps occur within 54~s.  Deadtime corrections have been incorporated into the spectra and data, and the Level-1 data excludes times when the count rates exceed the saturation limit. 
In some time bins, particularly during passes through polar radiation belts, DAXSS spectra include a particle background with an almost flat spectrum~\citep{2023ApJ...956...94W}. Average count rates in the 15--20~keV band, which do not generally include any significant contribution from solar X-rays, are used as the proxy for this flat particle background and are subtracted from the DAXSS spectra across all energies. Uncertainties on the count spectra are also obtained for each spectrum, considering Poisson statistics.    

We then generate XSM spectra for each of the DAXSS 54-s time bins using the XSMDAS Version~1.5~\citep{mithun20_soft} tool \texttt{xsmgenspec} from the Level-1 data of XSM\footnote{\url{https://pradan.issdc.gov.in/ch2/}}. Spectral generation excludes time bins when the Sun is not within XSM's field of view (FOV) or is occulted by the Moon, and it ignores time bins with gaps in the raw 1-second XSM data. As the XSM effective area varies with different Sun angles, corresponding effective area files (arf) are also generated along with the spectra. Deadtime corrections are incorporated in the spectra generated by correcting the exposure times; an update in the method of implementation of deadtime correction is given in Appendix~\ref{sec:xsm_deadtime}. Although the background is relatively low in XSM, we subtract the average non-solar background obtained from observations when the Sun is out of the FOV from the XSM spectra for each time bin. Uncertainties on the counts are also obtained for each spectra. Finally, we obtained simultaneous DAXSS and XSM solar X-ray spectra over 3671~time bins, corresponding to an effective exposure of $\sim$55~hours.

For comparisons of GOES-16 XRS flux with XSM, we use the Level-2 1-minute average measurements version~2.2.0\footnote{\url{https://www.ncei.noaa.gov/products/goes-r-extreme-ultraviolet-xray-irradiance}}. Although a higher cadence of 1~s is also available, we use the 1-minute average data to have sufficient counts in XSM spectra for estimating the flux. We use only the XRS flux measurements in channel-B (1--8~{\AA}) for the comparison, as this band is fully covered by XSM. The 1-minute average fluxes are corrected for electron contamination and have a lower threshold of $10^{-9} ~\mathrm{Wm^{-2}}$~\citep{2026JGRA..13135181M}. Good data (defined as those with the data quality flag set to zero) are selected for further analysis. XSM spectra are then generated for the corresponding 1-minute intervals, and time bins with data gaps are excluded. XSM count spectra are divided by the effective area, assuming diagonal response, and flux in the 1.55--12.4~keV band (1--8~{\AA}) is computed for each 1-minute bin. After filtering, we obtained $\sim$821,000 1-minute bins with valid GOES XRS and XSM fluxes, corresponding to an effective exposure of $\sim$570~days. 

\section{Results and Discussion}
\label{sec:result}

\subsection{Spectral Comparisons: XSM and DAXSS}

We first compare the count spectra obtained with XSM and DAXSS. It is very challenging to invert the observed count spectra from X-ray spectrometers into photon spectra in physical units, accounting for the spectral redistribution matrices and instrument effective areas. Hence, a forward-folding approach is used to model the X-ray spectra \citep[see the discussion in][]{mithun20_gcal}. 
As both XSM and DAXSS employ similar SDDs with similarly less prominent off-diagonal elements in the redistribution matrix, the count spectra can be divided by the respective effective areas to obtain spectra in unit area, assuming a diagonal response matrix. For each time bin, we divide the XSM and DAXSS spectra by their respective effective areas and compare them. However, the spectral resolutions of the two instruments are different. We convolve the DAXSS spectra with a Gaussian with energy-dependent width (estimated from the energy-dependent resolutions of the two instruments) such that the resolution matches that of XSM. A similar approach was used earlier by \citet{moore18} to compare MinXSS spectra with those from RHESSI.

\begin{figure*}
\begin{center}
   \includegraphics[width=0.99\textwidth]{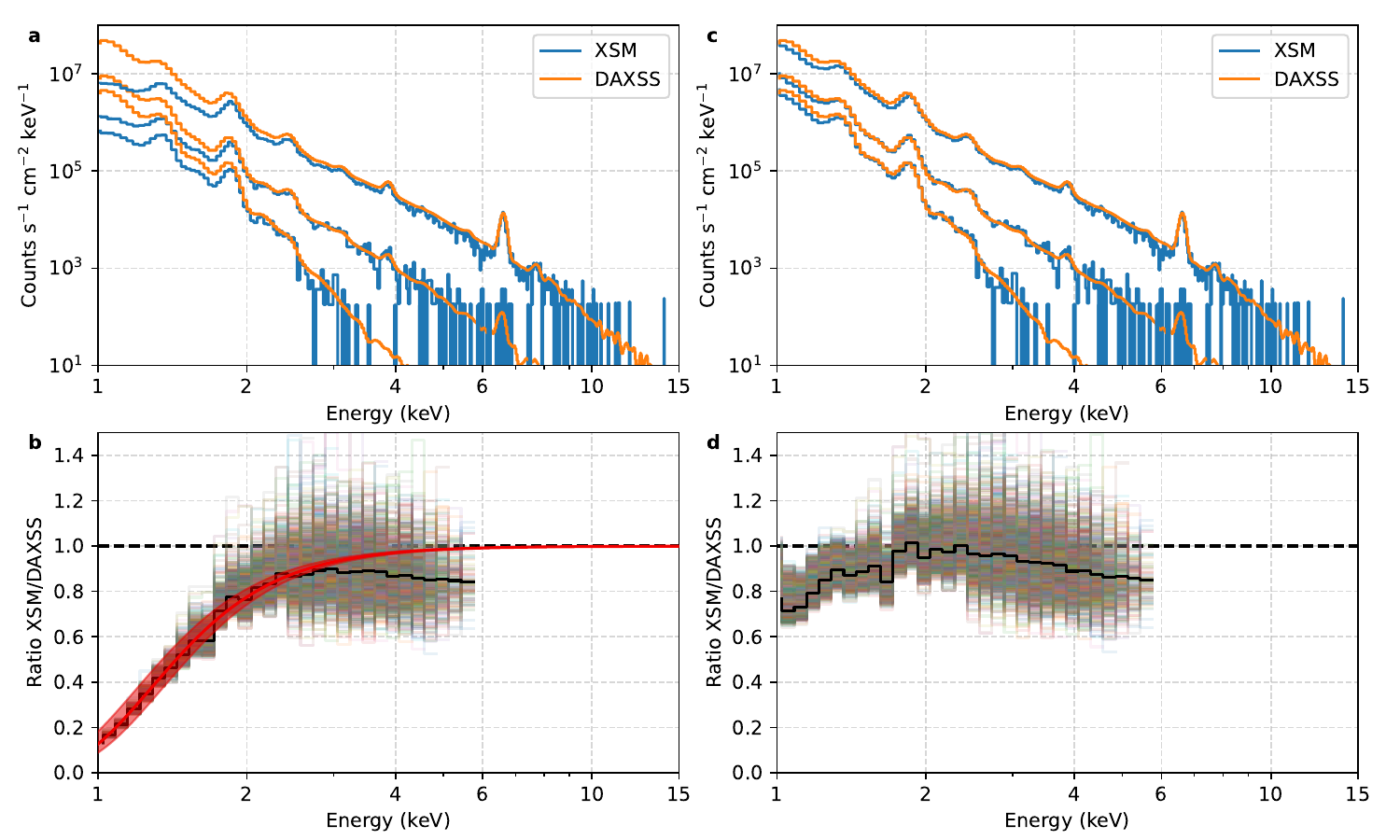}
   \caption{(a) Representative XSM and DAXSS spectra at three time intervals, scaled to unit area assuming diagonal response and matching the spectral resolutions (see text for details), where the current calibration of XSM is used. While the spectra match at high energies, clear differences are visible at lower energies. (b) The ratio of XSM spectra to DAXSS spectra, binned into broader energy bins, with different colors in the background for all available simultaneous observations. The average of the ratio is shown with a solid black line.  The overplotted red line corresponds to attenuation by \SI{17}{\micro\meter} of beryllium that is consistent with the observed ratio, and the red shaded region corresponds to the range of additional beryllium thickness from \SI{14}{\micro\meter} to \SI{20}{\micro\meter}. (c, d) Spectra and spectral ratios from (a) and (b), respectively, after incorporating the additional \SI{17}{\micro\meter} beryllium thickness in the XSM response, with better than 25\% agreement across energies between XSM and DAXSS. 
   \label{xsm_daxss_spec}}
\end{center}
\end{figure*}

Figure~\ref{xsm_daxss_spec}(a) shows XSM and DAXSS spectra obtained in this manner for three representative time bins. The figure shows that the spectra match reasonably well above $\sim$2~keV, whereas XSM systematically yields lower counts below this energy. To investigate this further, we bin the spectra into broader logarithmic energy bins and compute the XSM-to-DAXSS ratio for each time bin. Spectral bins with relative uncertainties greater than 40\% are not used to obtain the ratios. These spectral ratios are plotted as a function of energy in panel (b) of Figure~\ref{xsm_daxss_spec}. The average spectral ratio is shown as the black line in the figure. The decreasing trend of counts in XSM relative to DAXSS at lower energies, evident in the spectral overplot, is more pronounced in the ratio plot. 

The trend of the spectral ratio suggests a possible attenuation effect not accounted for in the XSM effective area or an overestimation of attenuation in the DAXSS effective area. Energy dependence of the DAXSS effective area at energies above $\sim$1.3~keV is primarily dictated by the transmission of a Kapton window placed in front of its detector. At even lower energies, the X-rays passing through the small clear aperture in the Kapton window are absorbed or transmitted by the beryllium window of the detector. The thickness of Kapton is measured physically, and the transmission at low energies, including that of the beryllium window, is obtained from measurements of DAXSS X-ray response at the National Institute of Standards and Technology (NIST)
Synchrotron Ultraviolet Radiation Facility (SURF)~\citep{2020ApJ...904...20S}.
Further, given the dual-zone nature of the DAXSS aperture, changes in the transmission of either beryllium or Kapton due to differences in thickness or attenuation coefficient cannot reproduce the observed monotonic trend in the spectral ratio with energy.  Therefore, we attribute the difference to unaccounted attenuation in the XSM effective area.

The energy dependence of the XSM effective area at low energies is determined by the transmission of its beryllium window, an integral part of the KETEK SDD module. While other aspects of the effective area, including the angular response of the detector, have been obtained from ground calibration experiments, attempts to measure the transmission of beryllium window with beamline experiments were not successful due to large uncertainties, and thus the thickness of the beryllium window is considered to be $8\substack {+5 \\ -0}$~\SI{}{\micro\meter} based on the specifications by the manufacturer for estimating the effective area~\citep{mithun20_gcal}. Using in-flight observations of the quiet Sun, it was found that the beryllium window thickness is non-uniform across the detector plane, and the relative variation in the window thickness was measured using the same observations, with an accuracy of a few percent~\citep{2020SoPh..295..139M}. However, it was not possible to constrain the absolute thickness of the beryllium window using these measurements. Moreover, based on XSM modeling of the spectral data, it was found that the response at energies below $\sim$1.3~keV was poorly modeled and not recommended for spectral fitting, although the exact reason was unknown at that time~\citep{2020SoPh..295..139M}. 

Thus, suspecting the possibility of increased thickness of the beryllium window in XSM, we compare the obtained spectral ratio with the attenuation of additional thickness of beryllium. The red line in Figure~\ref{xsm_daxss_spec}(b) corresponds to attenuation by an additional \SI{17}{\micro\meter} beryllium, and the red shaded region corresponds to the range for beryllium thickness from \SI{14}{\micro\meter} to \SI{20}{\micro\meter}. As the XSM observes the Sun at different angles, the beryllium window's effective thickness and transmission depend on the angle; thus, the average Sun angle during the specific observations is accounted for when plotting the transmission. It can be seen that the spectral ratio matches well with an additional beryllium thickness of \SI{17}{\micro\meter}, indicating that the detector's beryllium window is effectively \SI{25}{\micro\meter} rather than \SI{8}{\micro\meter} as originally modeled. 

Calibrations of MinXSS and DAXSS spectrometers and GOES XRS at NIST SURF indicate that a different thickness of the beryllium filter than the physical thickness is required to model the SURF measurements~\citep{caspi2015,2020ApJ...904...20S,2024JGRA..12932925W}. Analysis of the DAXSS spectra at SURF suggests that the modeled transmission of the beryllium window for DAXSS requires a $\sim$\SI{1.7}{\micro\meter} thicker window compared to the physical thickness of \SI{12.5}{\micro\meter}~\citep{2020ApJ...904...20S}. Calibration of GOES-R XRS instruments at SURF found that the effective model thickness of beryllium is $\sim$\SI{520}{\micro\meter} compared to the physical thickness of \SI{600}{\micro\meter} for channel A  and is $\sim$\SI{53}{\micro\meter} compared to the physical thickness of \SI{60}{\micro\meter} for channel B, which means $\sim$10\% systematically lower thicknesses are required in the model~\citep{2024JGRA..12932925W}. These show that the attenuation coefficients for beryllium~\citep{1993ADNDT..54..181H} can have systematic uncertainties of the order of $\sim$10\%, resulting in effective beryllium thickness differing from the physical thickness by similar magnitudes.

In the case of XSM, the difference between the effective thickness of \SI{25}{\micro\meter} and the physical thickness of \SI{8}{\micro\meter} is more than 200\%, compared to the 10\% differences observed in the DAXSS and XRS calibrations, suggesting that this difference cannot be fully attributed to coefficient uncertainties. Thus, it is possible that the window has a physical thickness greater than originally assumed. 
This is further supported by experiments with another instrument, the Alpha Particle X-ray Spectrometer (APXS) flown on the Chandrayaan-2 and Chandrayaan-3 rovers~\citep{2020CSci..118...53S,2020P&SS..18704923M,2024Natur.633..327V}, which used detectors from the same batch as those used in XSM.
Using the data obtained on the ground with two identical models of the APXS instrument employing two detectors from this batch, we find that the thicknesses of the beryllium windows of those two detectors are not the same, as discussed in detail in Appendix~\ref{sec:apxs}. The difference in thicknesses of the windows of these two detectors is $\sim$\SI{17}{\micro\meter}, similar to our finding here. This suggests that while some detectors in the batch had a window thickness of \SI{8}{\micro\meter}, others had a thickness of \SI{25}{\micro\meter}, consistent with our conclusion from the comparison of XSM and DAXSS spectra. It is also worth noting that \SI{25}{\micro\meter} is another standard beryllium window thickness offered by the manufacturer, which makes this very likely. We checked with the manufacturer about the possibility of a discrepancy in the window thickness, and they confirmed that they have no traceable records to ascertain the beryllium window thicknesses of the detector supplied in this batch.  

Based on this, we updated the XSM data analysis software and the associated calibration database to account for a base thickness of \SI{25}{\micro\meter} for the beryllium window. It may be noted that the relative variation of thickness of the beryllium window as reported in \citet{2020SoPh..295..139M} still remains valid and is included in the data processing. Panel (c) of Figure~\ref{xsm_daxss_spec} shows the XSM spectra with updated effective area overplotted with DAXSS spectra, and panel (d) shows the spectral ratio. It can be seen that the measurements from the two instruments now match closely, with differences within $\sim$25\% over the entire energy range. 

There are still some systematic patterns seen in the spectral ratio in Figure~\ref{xsm_daxss_spec}d, including differences at lower energies that might be reduced with increasing the beryllium window thickness of XSM further. However, we refrain from such a modification in the effective area for the following reasons. Since a difference of \SI{17}{\micro\meter} corresponds to a beryllium thickness of \SI{25}{\micro\meter}, which is a standard window thickness from the manufacturer, we use that value instead of attempting to fit the spectral ratio to determine a possible thickness. 
As noted earlier, DAXSS calibrations indicate that the beryllium window thickness is about 10\% higher than the physical thickness for thin windows, whereas GOES XRS calibrations indicate a 10\% lower value than the physical thickness for thicker windows. Considering this systematic pattern, the effective model thickness required for XSM may likely be a few microns higher than 25~\um. 
Comparison of the attenuation of beryllium of different thicknesses with the ratio of XSM to DAXSS spectra in Figure~\ref{xsm_daxss_spec}(b) shows that the uncertainty in beryllium window thickness is of the order of a few \SI{}{\micro\meter} (similar uncertainties are also associated with the estimates from the APXS instrument discussed in Appendix~\ref{sec:apxs}). 
As the uncertainties of obtaining the window thickness from the analysis are of a similar order as the expected difference due to attenuation coefficients, we use the mean window thickness of \SI{25}{\micro\meter}. 
It is also noted that an increase of thickness by 2--3~\SI{}{\micro\meter} would bring the ratio closer at energies below $\sim$1.5~keV, but this would cause an overshoot at energies around 1.8-- 2.5~keV where the corrected ratio is almost unity (see Figure~\ref{xsm_daxss_spec}(d)).
It is also worth noting that the current spectral-ratio comparison completely ignores spectral redistribution, i.e., the off-diagonal portions of the spectral response. While the impact is negligible compared to the larger differences with earlier effective areas, it becomes important to consider when the differences are of the order of 10\%. This effect is also seen in the spectral ratio in Figure~\ref{xsm_daxss_spec}(d), showing a step-like feature near the Si~K-edge. Based on these considerations, the update to the effective area of the XSM uses a base thickness of \SI{25}{\micro\meter} for Be. This update has been incorporated into XSMDAS Version~2.0, and the associated CALDB, and the further analysis in this paper uses this updated XSM calibration.

\subsection{Flux Comparison: XSM and DAXSS}

We now proceed to compare the flux measurements from XSM and DAXSS in broader energy bands. We choose two energy bands for this comparison: 1.55--12.4~keV, corresponding to the standard wavelength band 1--8~{\AA}, over which GOES instruments report solar flux, and 1--15~keV, which covers the entire common energy range of the two instruments. We begin with the XSM and DAXSS spectra in common time bins as discussed in the previous section and divide each by its respective effective area to obtain the spectra in flux units. Further, spectra in the two energy bands are integrated to obtain estimates of solar X-ray flux for both instruments.

\begin{figure}
\begin{center}
   \includegraphics[width=0.6\columnwidth]{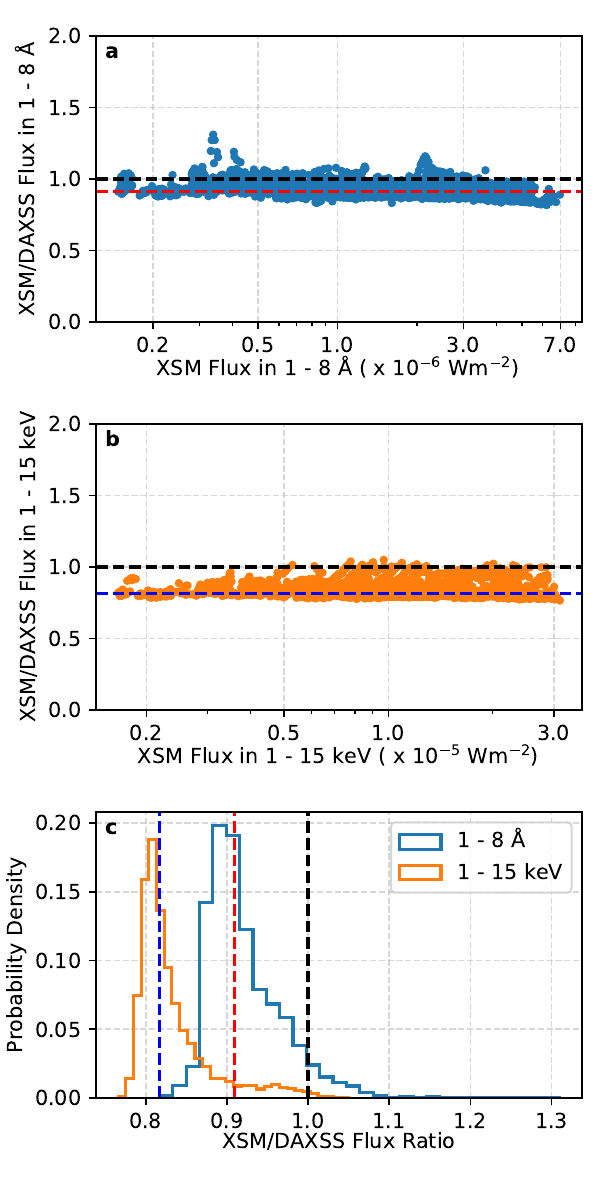}
   \caption{Ratio of XSM flux to DAXSS flux in 1--8~{\AA} (a) and 1--15~keV (b) bands is plotted as a function of XSM flux in the respective energy bands. Median values of the ratios of the fluxes are shown by the red and blue dashed lines in the respective panels. (c) Histograms of the ratios of the fluxes, showing that the fluxes agree within $\sim$10\% in the 1--8~{\AA} band and within $\sim$20\% in the 1--15~keV band. 
   \label{xsm_daxss_flux}}
\end{center}
\end{figure}

Figure~\ref{xsm_daxss_flux}(a) shows the ratio of XSM measurement of flux in 1--8~{\AA} to that of DAXSS plotted as a function of XSM flux. The median value of the ratio is~0.92, shown by the red dashed line in the figure. Similar flux ratio for the 1--15~keV band is shown in panel (b) of Figure~\ref{xsm_daxss_flux} and the median ratio of~0.82 is shown by the blue dashed line. Histograms of the ratio at each time bin for the two energy bands are shown in Figure~\ref{xsm_daxss_flux}(c). The distributions of the flux ratios are not very wide, except for a few outliers in which XSM fluxes are higher than DAXSS, visible in the figure as a cluster of flux values. Analysis of such rare cases shows that they correspond to a few flares, and the mismatch is most likely due to slight timing errors between the instruments. 
We also note a slight trend for XSM to underestimate flux relative to DAXSS at the highest flux values. This may be due to overcorrection of non-linearity in DAXSS flux that becomes important at those flux levels~\citep{2024JGRA..12932925W}.   
Overall, the 1--8~{\AA} flux measured by XSM and DAXSS agrees within $\sim$10\%, whereas the 1--15~keV fluxes agree within $\sim$20\%, with a slightly lower flux estimate from XSM compared to DAXSS. Given the typical absolute flux calibration uncertainties of about 10\%, this level of agreement indicates that the fluxes match within those uncertainties. 

\subsection{Flux Comparison: XSM and GOES XRS}

The comparisons between XSM and DAXSS presented in the previous sections were limited to a relatively narrower range of solar X-ray fluxes, given the limited opportunities for simultaneous observations. The GOES-16 XRS instrument (referred to as XRS hereafter) operated simultaneously with XSM for more than 5.5~years since the solar minimum period in 2019--2020. Although detailed spectral comparisons, such as those with DAXSS, are not feasible, broadband flux measurements with XSM can be compared with XRS over a much wider range of fluxes and time. As discussed in Section~\ref{sec:obs}, we select the 1-minute XRS channel B (1--8~{\AA}) fluxes and compute the XSM flux in the corresponding energy band (1.55--12.4~keV) for the respective time bins. XSM employs a \SI{250}{\micro\meter} beryllium attenuator in front of the detector during intense solar flares $\gtrsim$M6 class, attenuating counts below 2~keV to extend the dynamic range of measurements. Thus, in such cases, XSM measures the solar X-ray spectrum only from 2~keV onwards and cannot reliably provide flux measurements for the full 1--8~{\AA} band, and thus the flux comparisons are limited to $\lesssim$M6 class to avoid the attenuator.  

\begin{figure}
\begin{center}
   \includegraphics[width=0.6\columnwidth]{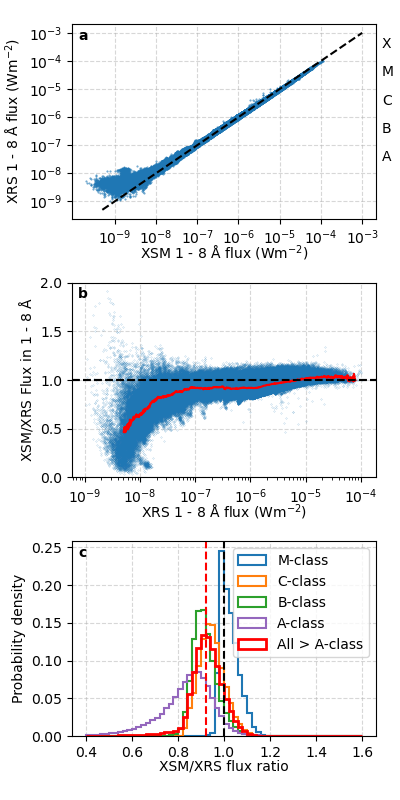}
   \caption{(a) Correlation of the XSM measured flux in 1--8~{\AA} and GOES-16 XRS Channel-B flux. Black dashed line shows one-to-one correlation for reference, and $y$-axis labels on the right side denote the classification of flux levels. (b) Ratio of the XSM measured flux to GOES XRS flux as a function of the GOES flux. Mean of the ratio in each flux bin is shown by the red line. (c) Histograms of the ratio of the fluxes for different flux levels and all fluxes above the A-class level. It can be seen that there is very good agreement between fluxes across a larger class of flares, and that this agreement systematically decreases for lower classes, as one would expect given the flat-solar-spectrum assumption in XRS flux estimation~\citep{2024JGRA..12932925W}. Overall, the agreement is within $\sim$20\% above A-class even without incorporating any correction for the flat spectrum assumption. 
   \label{xsm_goes}}
\end{center}
\end{figure}

Figure~\ref{xsm_goes}(a) shows the XSM flux plotted against XRS Channel-B flux, with the black dashed line showing the one-to-one correlation for reference. It can be seen that fluxes match well, especially at higher flux levels. The agreement is better examined by the ratio of the XSM flux to XRS flux in Figure~\ref{xsm_goes}(b) and the histogram of the ratio for different classes of flares in Figure~\ref{xsm_goes}(c). For convenience in Figure~\ref{xsm_goes}(c), the flux levels are defined by the XRS flare classes (A, B, C, M, X) where X is $\ge$1~$\times$~10$^{-4}$~Wm$^{-2}$, M~is 1~$\times$~10$^{-5}$ to 1~$\times$~10$^{-4}$~Wm$^{-2}$, etc. as shown in Figure~\ref{xsm_goes}(a). The mean of the ratio as a function of the XRS flux is shown by the red line in Figure~\ref{xsm_goes}(b). The ratio of the fluxes is closest to unity for the largest of the flares, and the XSM flux is slightly lower than the XRS flux for progressively less intense solar activity.  

The estimation of X-ray flux by the XRS instrument makes an assumption that the X-ray spectrum in the 1--8~{\AA} band is flat to convert the measured detector current to flux using XRS responsivity curves~\citep{2024JGRA..12932925W}. Because the flat spectrum assumption is not representative of the actual solar X-ray spectrum (see Figure~\ref{xsm_daxss_spec}), the XRS-reported flux is only an estimate of the solar flux, and the true flux can be different from this estimate. \citet{2024JGRA..12932925W} computed the potential differences between the actual flux and true XRS flux, considering the APEC models during active and quiet periods of solar activity. They reported that during active flaring, the XRS estimates are very close to the true fluxes, whereas during minimum periods the flux estimates can be about 2--3 times higher. This is consistent with the results we obtain for the fluxes measured by XSM compared to those measured by XRS. For the M-class flux level, the ratio of XSM flux to XRS flux is consistent with unity, and maximum departure is observed for lower classes of flux levels. 

It is possible to estimate the predicted XRS flux using measured flux from a spectroscopic instrument, taking into account the spectral shape as measured by the instrument, as done by \citet{2020ApJ...904...20S} for comparison of XRS flux with flux measured by DAXSS during a rocket flight. A similar comparison of INSPIRESat-1 DAXSS-predicted XRS flux with measured XRS flux shows that the DAXSS predictions are within 5\% for XRS channel~B for flux levels below C5~\citep{2026JGRA..13135181M}. However, in this work, we do not attempt such a comparison using predicted XRS fluxes and XSM observed spectra, as small changes in the effective area at low energies would lead to differences in the predicted fluxes. Even assuming a flat solar spectrum for XRS, the median flux ratio is~0.92 (8\%~difference), and flux estimates from XSM and XRS agree within 20\%. This agreement shows consistency between XSM and XRS over a wider range of flux levels over more than five years of observations.

\section{Summary}
\label{sec:summary}

Recent simultaneous observations of the Sun with several X-ray instruments have provided an opportunity to compare their measurements and cross-calibrate the instruments. Here, we presented the cross-calibration of the X-ray spectrometer Chandrayaan-2 XSM with the DAXSS spectrometer on board INSPIRESat-1 and the broadband instrument XRS on GOES-16, using strictly simultaneous solar observations. Comparisons of the X-ray count spectra of XSM and DAXSS showed the presence of an attenuation effect unaccounted for in the XSM effective area. Based on the spectral ratio of XSM to DAXSS and other laboratory measurement evidence, the difference is attributed to the thickness of the XSM detector's beryllium window being \SI{25}{\micro\meter} thick instead of \SI{8}{\micro\meter}. With the updated effective area of the XSM, which is also included in the version~2.0 release of the XSM Data Analysis Software, the XSM spectra and DAXSS spectra agree within 25\% over the range of energies, with much better agreement over some energy bands. Flux measurements with XSM and DAXSS agree within $\sim$20\% in the 1--15~keV and within $\sim$10\% in 1--8~{\AA} (1.55--12.4~keV). Comparison of XSM measured flux with GOES-16 XRS channel~B flux shows a very good agreement at higher flux values and progressively slightly lower flux in XSM at lower fluxes, consistent with the flat spectrum assumption used in XRS flux estimate. Even with this flat spectrum assumption, the median difference between XRS flux and XSM flux is less than 10\%.  

This work has resolved previously unknown response issues in XSM, most evident at energies below 1.3~keV. With the updated calibration, XSM spectroscopy is now possible from 1~keV onwards, compared to the earlier recommended lower energy limit of 1.3~keV. With the updated calibration, median differences in flux measured by XSM with DAXSS and XRS are within $\sim$10\%. In this work, we have employed a spectral model-independent approach to compare the spectra observed by XSM and DAXSS, using a diagonal spectral redistribution matrix. A comparison accounting for this would be possible by spectral modeling of XSM and DAXSS, and comparing the derived spectral parameters, which would be addressed in a follow-up work.

\backmatter

\bmhead{Acknowledgements}

We acknowledge the use of data from the Solar X-ray Monitor (XSM) on board the Chandrayaan-2 mission of the Indian Space Research Organization (ISRO), archived at the Indian Space Science Data Center (ISSDC). XSM was developed by Physical Research Laboratory (PRL), Ahmedabad, India, with support from various ISRO centers. 
We also acknowledge the use of data from Dual-zone Aperture X-ray Solar Spectrometer (DAXSS) developed by LASP on board INSPIRESat-1. INSPIRESat-1 was developed by the University of Colorado, Boulder, USA; the Indian Institute of Space Science and Technology, Thiruvananthapuram, India; the Nanyang Technological University, Singapore; and the National Central University, Taiwan, and was launched by ISRO.
This work utilizes data from the GOES-16 X-ray Sensor (XRS) provided by the National Oceanic and Atmospheric Administration (NOAA).
The research work at Physical Research Laboratory, Ahmedabad, is supported by the Department of Space, Government of India.

\bmhead{Data availability} Chadrayaan-2 XSM data are available at the PRADAN portal of Indian Space Science Data Archive (ISAD) at \url{https://pradan.issdc.gov.in/ch2/}, INSPIRESat-1 DAXSS data are available at \url{https://lasp.colorado.edu/minxss/data/}, and GOES-16 XRS data are available at \url{https://www.ncei.noaa.gov/products/goes-r-extreme-ultraviolet-xray-irradiance}.

\section*{Declarations}

\bmhead{Competing interests} The authors declare no competing interests.

\begin{appendices}

\section{Deadtime corrections for XSM}
\label{sec:xsm_deadtime}

Based on the validation of the deadtime model using ground calibration experiments, a method was devised to correct the deadtime effects in the observed spectrum~\citep{mithun20_gcal}.
This is done by modifying the exposure time $(T_{\mathrm{exp}})$ of the observed spectrum by:

\begin{equation}    
T_{\mathrm{live}} = T_{\mathrm{exp}} * \frac{n_{\mathrm{d}}}{n_{\mathrm{t}}} * (1~-~n_{\mathrm{t}}~{\tau}_{\mathrm{1}})
\label{deadCorEqn1}
\end{equation}

\noindent where $n_{\mathrm{t}}$ and $n_{\mathrm{d}}$ are the trigger and detected rates, respectively, and ${\tau}_{\mathrm{1}}$ is the deadtime corresponding to pulse shaping time of \SI{0.96}{\micro\second}.

However, on analysis of in-flight observations with XSM, it was observed that the trigger rates were typically 20--50 counts higher than the detected count rate, even when the average count rate was as low as 0.15~counts/s during background observations (excluding ULD events of $\sim$1~counts/s). Furthermore, it is noted that these excess triggers are observed when ULD (Upper-Level Discriminator) events occur in the same second. Based on this, we infer that these additional trigger events are most likely the result of the ringing effect of the pulse after a particle-induced event, which usually gets recorded as a ULD event. As these follow immediately after the ULD event, they do not significantly affect the deadtime of the instrument.

\begin{figure}[h!]
\begin{center}
    \includegraphics[width=0.7\columnwidth]{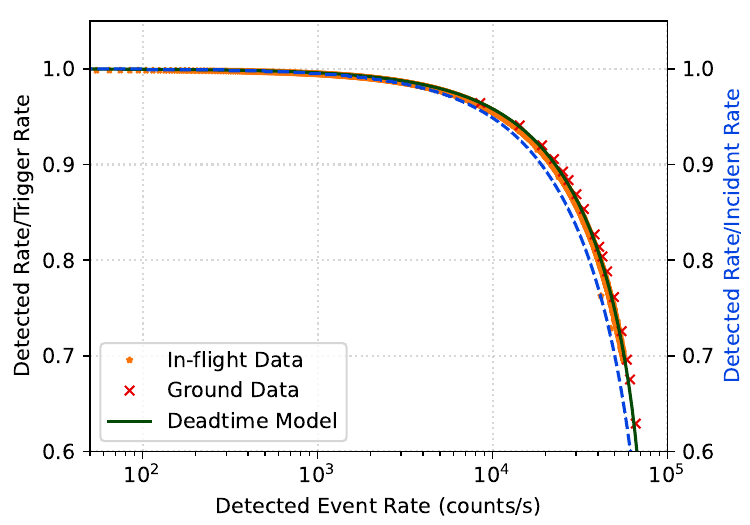}
    \caption{Ratio of event triggers to detected events plotted as a function of detected event rates. 
    \label{deadtimeCor}}
\end{center}
\end{figure}

In Figure~\ref{deadtimeCor}, the observed ratio of detected events to trigger rates is plotted against detected event rates, averaged over the times when ULD events are not observed. Ratio of detected rates to event trigger rates obtained from ground calibration experiment, as discussed in \citet{mithun20_gcal}, is overplotted in red. The prediction from the deadtime model is overplotted in green, which matches very well with the observed in-flight and on-ground data, indicating that the deadtime model can be used to correctly obtain the incident flux.   

As there are spurious events present in the trigger count, it would not be possible to use trigger rates to correct for deadtime effects using Equation~\ref{deadCorEqn1}. Instead, we use the following relation between detected rate ($n_{\mathrm{d}}$) and actual incident rate ($n_{\mathrm{a}}$):
\begin{equation}
n_{\mathrm{d}} = n_{\mathrm{a}} ~\mathrm{exp}(-n_{\mathrm{a}} {\tau}_{\mathrm{2}})
\end{equation}
\noindent where ${\tau}_{\mathrm{2}}$ is the paralyzable deadtime of \SI{5}{\micro\second} implemented in XSM. 
However, this cannot be inverted to compute the actual incident rate from the detected rate. Instead, we generate a lookup table between $n_{\mathrm{d}}$ and $n_{\mathrm{a}}$ based on this relation and use that to get the incident rate corresponding to the detected rate, the ratio of which is also shown in Figure~\ref{deadtimeCor}. From the estimated incident rate based on the lookup table and the detected rate, the exposure times are modified as:

\begin{equation}    
T_{\mathrm{live}} = T_{\mathrm{exp}} * \frac{n_{\mathrm{d}}}{n_{\mathrm{a}}}
\label{deadCorEqn2}
\end{equation}

\noindent Deadtime corrections following this method have been incorporated in the XSM Data Analysis Software (XSMDAS) Version~1.5 onwards and are included in the analysis presented in this paper. 

\section{Window thickness differences of detectors: Measurements with APXS instrument}
\label{sec:apxs}
The Alpha Particle X-ray Spectrometer (APXS) instrument flown on the rovers of Chandrayaan-2 and Chandrayaan-3 missions irradiates the target sample with alpha particles and X-rays from its $^{244}$Cm source and records the fluorescent X-ray emission for in-situ qualitative and quantitative elemental analysis~\citep{2020CSci..118...53S,2020P&SS..18704923M,2024Natur.633..327V}.
APXS also employed an SDD similar to that in XSM. The SDDs used in XSM, as well as different models of the APXS instrument (flight models for the two missions and a spare model), were all procured from the manufacturer at the same time. For the intended purpose of the APXS instrument to measure the elemental abundances of lunar soil, knowledge of the exact detector window thickness was not essential, as an empirical calibration method was employed using the measured fluorescence spectra of standard geochemical reference materials~\citep{2020P&SS..18704923M}. However, for calibration, fluorescent X-ray spectra from various standard samples were acquired with different models of the APXS instrument in identical geometric configurations \citep[see][ for details of the experiment setup]{2020P&SS..18704923M}, and these could be used to see if there were any differences in the properties of different detectors in the same batch.

\begin{figure}
\begin{center}
   \includegraphics[width=0.99\columnwidth]{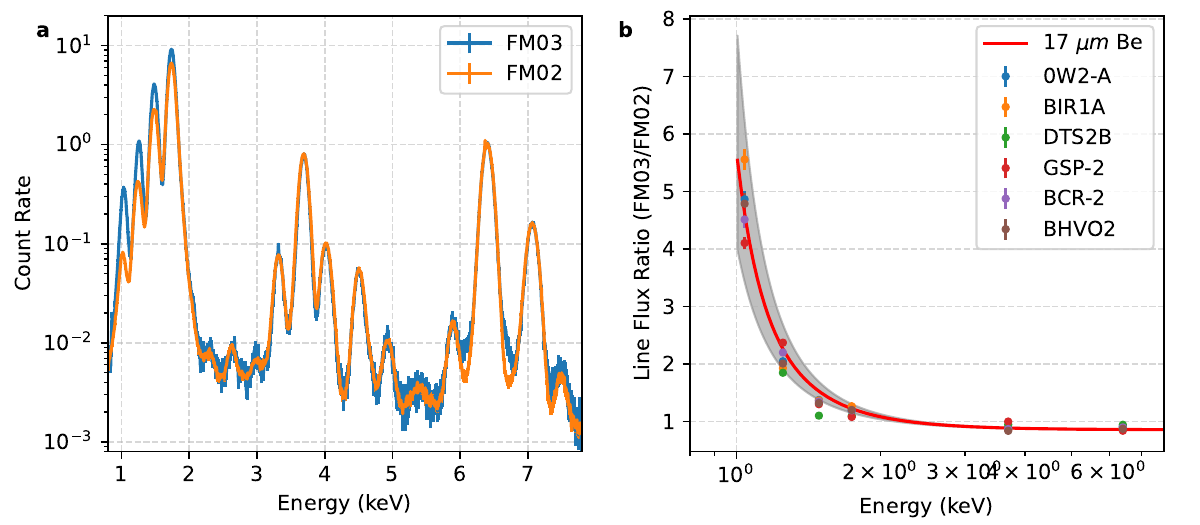}
   \caption{(a) APXS measured spectra of one of the geochemical reference materials at identical observing conditions with FM02 and FM03 models of the instrument. FM03 shows higher fluxes for the low-energy lines.  (b) Ratio of fitted line fluxes from FM03 and FM02 models obtained with different geochemical reference material spectra. The red line corresponds to the difference expected from an extra \SI{17}{\micro\meter} thickness for the beryllium window in the detector for FM02 compared to that of FM03, and the grey shaded region corresponds to the range for beryllium thickness from \SI{14}{\micro\meter} to \SI{20}{\micro\meter}.
   \label{apxsThick}}
\end{center}
\end{figure}

Figure~\ref{apxsThick}, left panel, shows two sets of fluorescence spectra of a geochemical sample acquired by FM02 and FM03 models of the APXS instrument. The figure shows that the lines at higher energies (beyond 3~keV) exhibit very similar fluxes in both instruments, whereas those at lower energies show a systematic difference. The FM03 instrument measures progressively higher fluxes compared to the FM02 instrument for the low-energy lines. To investigate this further, using spectral fits, we computed the line-flux ratio measured by the two instruments for different geochemical reference materials, which is plotted as a function of energy in the right panel of Figure~\ref{apxsThick}. A systematic trend in the ratio is clearly visible and consistent with a difference in window transmission between the two instruments, with the FM02 instrument having an additional thickness of \SI{17}{\micro\meter} compared to FM03 (red line overplotted in the figure). It may be noted that the strength of the radioactive source used for the excitation of fluorescence lines has also slightly reduced between these two observations and is taken into account in the overplotted red line (hence, slightly below one even at higher energies). This would mean that if the FM03 detector window has \SI{8}{\micro\meter} thickness, the thickness of the FM02 detector is \SI{25}{\micro\meter}, showing that at least some detectors in the batch have a window thickness of \SI{25}{\micro\meter} instead of \SI{8}{\micro\meter}. Thus, this analysis is consistent with the conclusion of a \SI{25}{\micro\meter}-thick beryllium window for the XSM instrument.  

\end{appendices}

\bibliography{references}

\end{document}